\documentclass[conference]{IEEEtran}
\IEEEoverridecommandlockouts
\usepackage{cite}
\usepackage{booktabs}
\usepackage{amsmath,amssymb,amsfonts}
\usepackage{algorithmic}
\usepackage{graphicx}
\usepackage{textcomp}
\usepackage{xcolor}
\usepackage{hyperref}
\usepackage{float} 
\def\BibTeX{{\rm B\kern-.05em{\sc i\kern-.025em b}\kern-.08em
    T\kern-.1667em\lower.7ex\hbox{E}\kern-.125emX}}

\begin{document}

\title{An Agentic Operationalization of DISARM for FIMI Investigation on Social Media}


    
    

\makeatletter
\newcommand{\linebreakand}{%
  \end{@IEEEauthorhalign}
  \hfill\mbox{}\par
  \mbox{}\hfill\begin{@IEEEauthorhalign}
}
\makeatother

\author{
    \IEEEauthorblockN{Kevin Tseng}
    \IEEEauthorblockA{\textit{Center of Research Acquisition} \\
    \textit{National Institute of Cyber Security}\\
    Taipei City, Taiwan \\
    kevin.tseng@nics.nat.gov.tw}
    \and
    \IEEEauthorblockN{Juan Carlos Fernandez Toledano}
    \IEEEauthorblockA{\textit{Department of Mathematics} \\
    \textit{Royal Military Academy}\\
    Brussels, Belgium \\
    juan.fernandeztoledano@mil.be}
    \and
    \IEEEauthorblockN{Bart De Clerck}
    \IEEEauthorblockA{\textit{Department of Mathematics} \\
    \textit{Royal Military Academy}\\
    Brussels, Belgium \\
    bart.declerck@rma.ac.be}
    
    \linebreakand 
    
    \IEEEauthorblockN{Yuliia Dukach}
    \IEEEauthorblockA{\textit{Research Division} \\
    \textit{OpenMinds Ltd.}\\
    Kyiv, Ukraine \\
    yuliia.dukach@openminds.ltd}
    \and
    \IEEEauthorblockN{Phil Tinn}
    \IEEEauthorblockA{\textit{Department of Sustainable } \\
    \textit{Communication Technologies} \\
    \textit{SINTEF}\\
    Oslo, Norway \\
    phil.tinn@sintef.no \\
    (Corresponding Author)}
}

\maketitle

\begin{abstract}
Interoperable data and intelligence flows among allied partners and operational end‑users remain essential to NATO’s collective defense across both conventional and hybrid threat environments. Foreign Information Manipulation and Interference (FIMI) increasingly spans multiple societal domains and information ecosystems, complicating threat characterization, persistent situational awareness, and coordinated response. Concurrent advances in AI have further lowered the barrier to conducting large‑scale, AI‑augmented FIMI activities—including automated generation, personalization, and amplification of manipulative content. While frameworks such as DISARM offer a standardized analytical and metadata schema for characterizing FIMI incidents, their practical application for automating large‑scale detection remains challenging. We present a framework‑agnostic, agent‑based operationalization of DISARM piloted to support FIMI investigation on social platforms. Our agent coordination pipeline integrates general agentic AI components that (1) identify candidate manipulative behaviors in social‑media data and (2) map these behaviors to DISARM taxonomies through transparent, auditable reasoning steps. Evaluation on two practitioner‑annotated, real‑world datasets demonstrates that our approach can effectively scale analytic workflows that are currently manual, time‑intensive, and interpretation‑heavy. Notably, the experiment surfaced more than 30 previously undetected Russian bot accounts—deployed for the 2025 election in Moldova—during the prior non-agentic investigation. By enhancing analytic throughput, interoperability, and explainability, the proposed approach provides a direct contribution to defense policy and planning needs for improved situational awareness, cross‑partner data integration, and rapid assessment of information‑environment threats.

This paper was originally presented at the International Conference on Military Communication and Information Systems (ICMCIS), organized by the Information Systems Technology (IST) Scientific and Technical Committee, IST-224-RSY—the ICMCIS, held in Bath, United Kingdom, 12-13 May 2026

\end{abstract}


\section{Introduction}
Foreign Information Manipulation and Interference (FIMI) is the term adopted by the European Union's External Action Service (EEAS) to describe state-sponsored disinformation and influence campaigns. Unlike generic “fake news,” FIMI is characterized by coordinated malign behaviors, often orchestrated by foreign actors with the intention of harming or destabilizing societies. Hostile actors are increasingly using artificial intelligence (AI) to supercharge their influence efforts. In addition to bots that amplify messages at inhuman speeds, generative AI provides a low-cost option\cite{Olejnik2026} for creating inauthentic content and increasing the scale of FIMI activities, accounting for 27\% of incidents in 2025 analyzed by EEAS \cite{EEAS2026Report}.

The Disinformation Analysis and Risk Management (DISARM) framework is an open-source taxonomy designed to standardize the identification and analysis of influence operations. The framework organizes threat actor behaviors into distinct tactics, techniques, and procedures (TTPs), creating a common vocabulary for researchers, policymakers, and defenders. By systematically mapping the operational steps of a campaign, it facilitates cross-sector data sharing, enables more precise attribution of adversarial behaviors, and supports the development of effective, evidence-based countermeasures against FIMI \cite{9830669}.


Despite the analytical value of DISARM, operationalizing it at social media scale remains difficult. Mapping TTPs typically requires labor-intensive, expert-driven coding of large, multilingual, and rapidly evolving datasets. This creates a gap between the conceptual promise of the framework and its day-to-day usability for timely detection, attribution, and response.

Agentic AI represents a leap in artificial intelligence, evolving beyond static, rule-based tools into autonomous systems capable of pursuing complex, open-ended goals with minimal human intervention. Characterized by advanced reasoning, adaptive decision-making, and the ability to decompose high-level objectives into executable sub-tasks, Agentic AI can operate effectively in dynamic and unstructured environments \cite{10849561}.

In this work, we propose an agent-based operationalization of DISARM to investigate FIMI on social media. We develop an agent coordination pipeline in which publicly accessible agentic AI components collaboratively (1) detect candidate manipulative behaviors and (2) map these behaviors onto DISARM TTPs in a transparent manner.

We evaluate the approach on two real-world datasets annotated by FIMI domain practitioners. This allows us to test whether autonomous agents can detect known TTP patterns, how robustly they generalize across platforms and topics, and where human oversight remains essential.
We make three contributions to the practice of FIMI investigation:

\begin{enumerate}
    \item \textbf{Technique-guided agent architecture}: We present a modular agent architecture that operationalizes DISARM as an executable investigative workflow, featuring technique-guided hypothesis generation, in which every discovered pattern maps to a specific TTP, and subprocess relay, which enables consistent execution across 15 investigation rounds.

    \item \textbf{Verifiable evidence pipeline}: We introduce atomic evidence decomposition—breaking complex findings into discrete, verifiable claims—combined with statistical validation, providing reproducible \texttt{PASS/FAIL} criteria for LLM-generated evidence.

    \item \textbf{Military application analysis}: We discuss how this agent-based DISARM operationalization can enhance situational awareness, improve interoperability through standardized outputs, and enable effective human-AI teaming in real-world FIMI defense.
\end{enumerate}

\section{Rationale to military application}
For military and defense organizations, the information environment is an operational space within the Multi-Domain Operations (MDO) paradigm, closely intertwined with, but not reducible to, the cyber domain. FIMI is increasingly treated as a key instrument of hybrid warfare, in which information manipulation is used alongside other instruments of power to achieve strategic effects \cite{siman2022hybrid}. Because such campaigns can generate strategic effects below the threshold of armed conflict and at high tempo, defensive FIMI investigation must be scalable, structured, and aligned with alliance practices in order to support timely and proportionate responses. From a doctrinal perspective, FIMI defense maps naturally onto the classical OODA loop (Observe–Orient–Decide–Act). Rapid observation of emerging narratives, robust orientation through structured interpretation, timely decision on proportional responses, and coordinated action across stakeholders are prerequisites for maintaining initiative in the cognitive domain. Agentic AI provides a mechanism to compress the observe–orient phases by automating continuous analysis\footnote{We do not focus on agentic collection in this work}, while DISARM supplies a stable semantic backbone to ensure that machine-produced outputs remain interpretable and operationally meaningful \cite{nadibaidze2024ai}. An agent-based operationalization of DISARM therefore offers a concrete pathway to integrate FIMI defense into Command, Control, Intelligence, Surveillance, and Reconnaissance (C2ISR) processes.

\subsection{Situational Awareness}
Military decision-making depends on shared situational awareness, increasingly understood as multi-domain awareness that includes the information environment \cite{JCS2018JCOIE, NATOACT2025MDO}. In FIMI contexts, awareness requires persistent monitoring of social media networks at scale, rapid detection of coordinated inauthentic behavior (CIB), and mapping of narrative spread across platforms and languages \cite{kalensky2025countering}. The speed and volume of AI-supported influence operations mean that delays in detection can directly translate into strategic disadvantage. An agentic DISARM supports situational awareness by allowing autonomous agents to collect signals, identify candidate manipulative behaviors, and continuously map these behaviors to DISARM TTPs. This produces structured, time-sensitive indicators of adversarial activity rather than unorganized streams of alerts. In practice, it allows human analysts and commanders to move from ad hoc monitoring to an always-on early-warning posture, improving both the timeliness and the granularity of FIMI threat pictures. Importantly, this awareness is actionable because it is expressed in a standardized behavioral taxonomy, facilitating prioritization and response planning.

\subsection{Human-Machine Teaming}
A recent CSIS study \cite{JensenStrohmeyer2025} on agentic warfare argues that Napoleonic‑style, centralized staffs are too slow and brittle for conflicts in which AI agents compress decision cycles from days to minutes and adversaries deliberately target command networks—thereby necessitating smaller, adaptive, AI‑enabled staffs that preserve human judgment while operating at machine speed. In this context, our proposed workflow treats agentic systems as accelerators of staff work rather than replacements for FIMI analysts and responders: agents continuously monitor, decompose tasks, surface hypotheses, and organize evidence, while expert analysts validate, contextualize, and interpret \cite{CEwithAI}—especially in inherently interpretive FIMI workflows that require assessing intent, cultural cues, and proportionality, where full automation is neither desirable nor feasible. Operationalizing DISARM with agents provides the doctrinal scaffold analogous to CSIS recommendations for resilient human–machine teaming: instead of opaque labels, agents must justify each assignment as an instance of specific TTPs within an explicit taxonomy, yielding explainable, traceable outputs from raw content to analytic conclusions, reducing automation bias, and enabling auditability. In short, autonomous agents extend analyst capacity at machine speed, while DISARM‑aligned outputs maintain transparency, accountable judgment, and doctrinal coherence—exactly the combination required for the adaptive staff models \cite{JensenStrohmeyer2025} envisioned for future military operations.

\subsection{Interoperability}
In alignment with NATO’s strategic move toward a "digitally enabled, multi-domain alliance," an agentic DISARM pipeline serves as a critical "interoperability-by-design" solution that addresses the \textit{technical}, \textit{procedural}, and \textit{human} dimensions of collective defense \cite{NATOACT2025Interoperability}. By grounding itself on standardized frameworks such as DISARM or STIX\cite{OASISSTIX21} as an ontological foundation, the pipeline establishes a \textit{procedural} common language that standardizes investigative outputs across national contexts, effectively reducing the "interpretive drift" that often hinders joint responses to foreign influence. This semantic standardization enables \textit{technical} interoperability through the multi-layer fusion \cite{10.1007/s10489-016-0834-7, 10.1016/j.inffus.2007.01.004} of incident data. From a \textit{human} interoperability perspective, the efficiency gain through data fusion, in turn, facilitates the emergence of an agentic military staff \cite{JensenStrohmeyer2025} able to process information at a speed and across scales that can exceed those of traditional staffing structures. Ultimately, by automating the cross-validation and attribution of transnational threats, the "agent-augmented" pipeline can transform fragmented national-level analyses into a unified, shareable ecosystem, enabling collective defense against transnational information threats.

\section{Methods}
This section presents our agent-based pipeline for operationalizing the DISARM framework, developed for FIMI investigations. The pipeline, built on BMAD-METHOD (Breakthrough Method for Agile AI-Driven Development)\cite{BMADMethod}, enables systematic discovery and statistical verification of information operations (IO) indicators through iterative hypothesis generation and testing.

\subsection{Pipeline overview}
Our approach addresses three key challenges in automating FIMI detection: (1) mapping discovered anomalies to standardized IO taxonomies, (2) ensuring statistical rigor in evidence validation, and (3) maintaining full traceability from raw data to verified findings.
We use “agentic” to refer to a workflow centered on a single LLM that runs across multiple investigation iterations; this does not imply a classical multi-agent system of independent agents. The current implementation operates on textual social media data—posts, comments, and associated metadata (e.g., timestamps, account identifiers, etc.)—though the framework is modality-agnostic by design. The pipeline operates in two phases, as illustrated in the upper and lower parts of Figure \ref{fig:pipeline}:

\begin{enumerate}
    \item \textbf{Investigation planning and execution}: An LLM agent iteratively explores the dataset, selecting DISARM techniques, generating hypotheses about potential IO indicators, and testing them with quantitative measures.
    \item \textbf{Evidence assessment}: Complex findings are decomposed into discrete, verifiable atomic evidence units; each atomic evidence claim is converted to a detection condition and statistically validated against ground truth labels or any available annotations.
\end{enumerate}

\noindent In implementing the pipeline, we considered the following design principles:
\begin{itemize}
    \item \textbf{Deferred anomaly detection}: The initial exploration phase focuses on establishing a neutral understanding of the structure, scale, and behavioral baselines without attempting to identify anomalies or assign interpretive meaning. This separation prevents premature, framework-driven interpretation and allows DISARM mapping to emerge only during later, technique-guided iterations, when hypotheses are explicitly formulated and tested.
    \item \textbf{Full history feedback}: Each iteration accesses the complete history of prior findings, enabling informed technique selection and contextual reasoning across iterations.
    \item \textbf{Evidence-based verification}: All claims undergo statistical validation using ground-truth labels or annotations, with explicit \texttt{PASS/FAIL} criteria based on effect-size thresholds.
    \item \textbf{Natural language specification}: Workflows, tasks, and agent behaviors are defined using natural language rather than traditional code, enabling rapid adaptation to diverse OSINT scenarios without requiring software development expertise.
\end{itemize}

\subsection{Dataset overview}

\subsubsection{Chinese campaign on X focused on the coordinated attacks on Guo Wengui in 2017/2019}
We selected the \texttt{{China~1}} and \texttt{China~2} datasets introduced by Seckin \cite{seckin2024} from a large collection of labeled information operations (IO) datasets. These datasets originate from transparency disclosures on a social media platform and correspond to two distinct but thematically related IO campaigns attributed to Chinese state-backed actors. Figure~\ref{fig:china_dataset} illustrates representative posts and a summary of the data set.

\begin{figure}[t]
    \centering
    \includegraphics[width=\columnwidth]{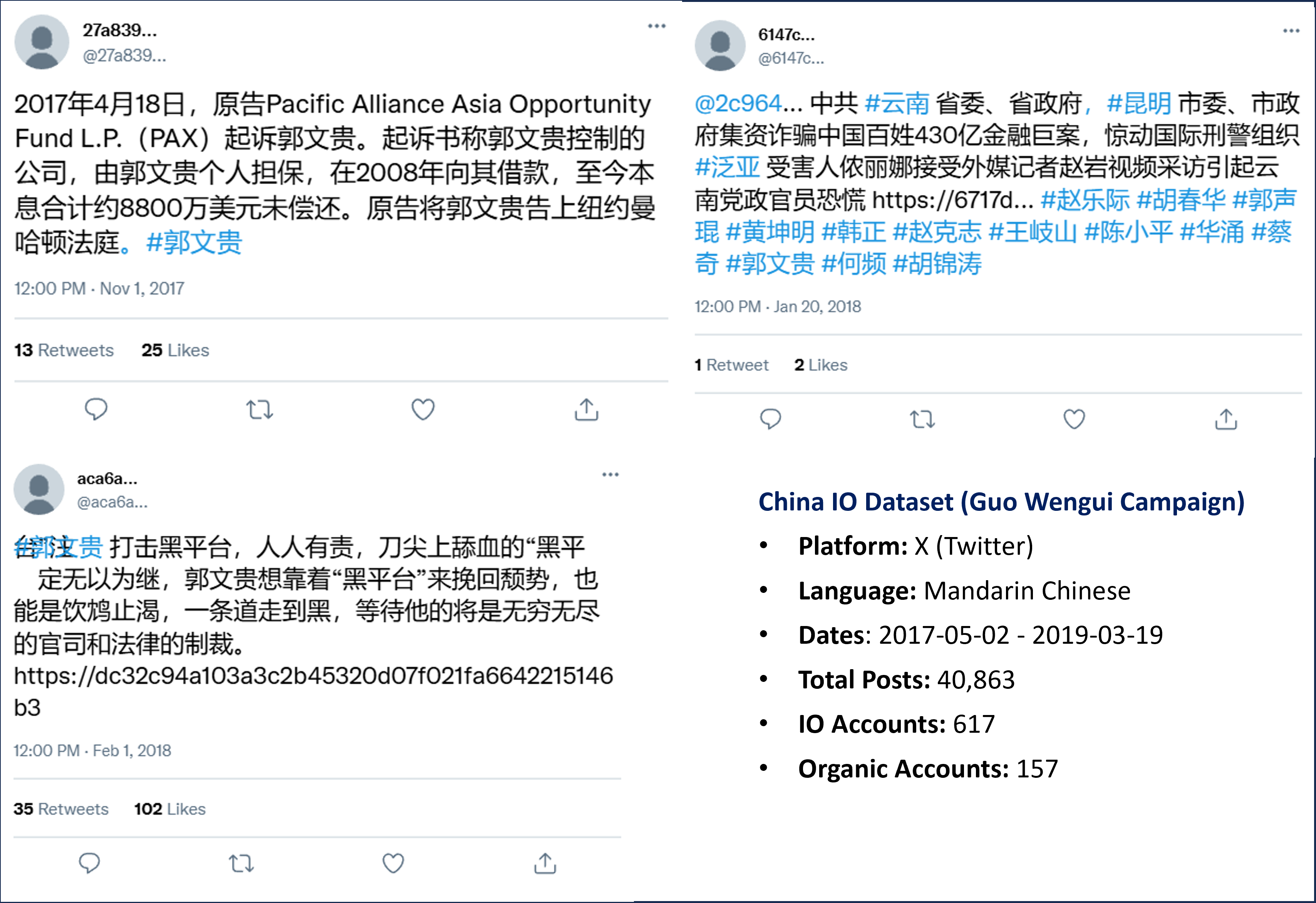}
    \caption{Sample X posts(in original Simplified Mandarin Chinese language) and core metrics of the China IO campaign against Guo~Wengui.}
    \label{fig:china_dataset}
\end{figure}

In the original datasets, the platform explicitly identified and verified that the IO accounts participated in coordinated information operations. To enable supervised evaluation of IO detection methods, the authors complemented the IO data with a control dataset representing organic user activity. Control accounts are defined as accounts not identified as IO by the platform but that discuss similar topics during the same time periods as the IO campaigns, ensuring topical and temporal overlap between IO and non-IO activities. Control accounts are collected using a hashtag-based matching strategy: hashtags used by IO accounts on specific dates are used to identify other accounts posting on the same dates with at least one of those hashtags.

In this work, we merge \texttt{{China~1}} and \texttt{China~2} into a single dataset. We then apply additional processing steps to obtain a simplified dataset focused on coordinated attacks on Guo~Wengui, a Chinese businessman and political dissident whose activities and public statements have made him a recurrent target of state-aligned online campaigns. First, we restrict the data to posts in Mandarin Chinese related to the campaign using a curated list of keywords and hashtags associated with narratives and attacks targeting Guo~Wengui. This focuses on the central content of the information operation while preserving both IO and control users. Second, we apply a uniform text-cleaning pipeline to both IO and control posts, including the removal of URLs, mentions, hashtags, emojis, and extraneous whitespace, while preserving Chinese characters. The posts are tokenized, and only posts with at least 10 valid tokens are retained to eliminate short or noisy messages.

The resulting dataset consists of 617 IO accounts and 157 organic accounts, comprising 40{,}863 cleaned Chinese-language posts (38{,}927 IO posts and 1{,}936 organic posts) drawn from both platform-verified IO accounts and organic control accounts, sharing a common topical and temporal context.

\begin{figure}[t]
    \centering
    \includegraphics[width=\columnwidth]{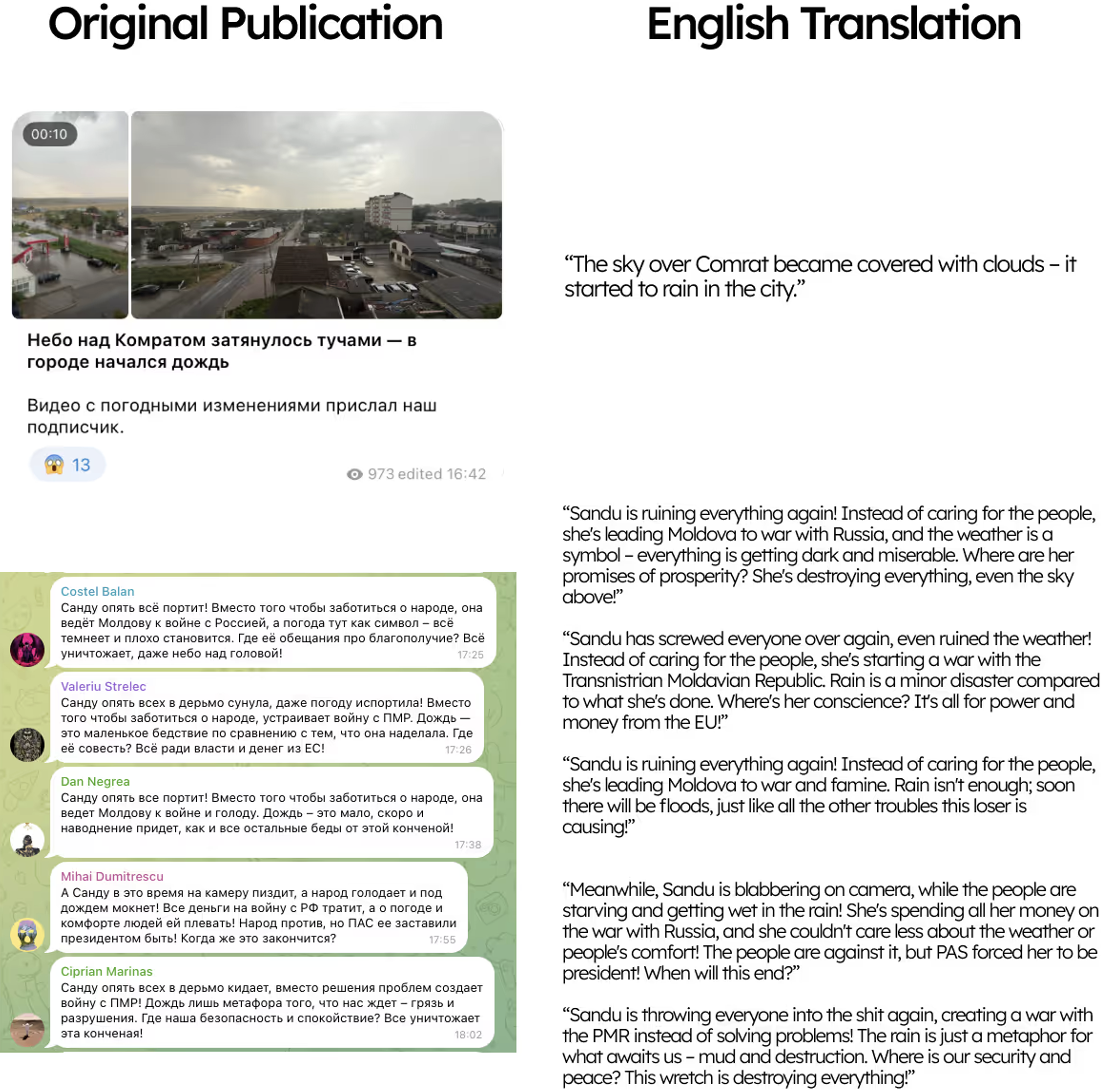}
    \caption{Sample text of glitched Telegram comments generated by Russian bots attempting to link the rain to discrediting the targeted politician Maia Sandu.}
    \label{fig:moldova_dataset}
\end{figure}

\begin{figure*}[t]
    \centering
    \includegraphics[width=\textwidth]{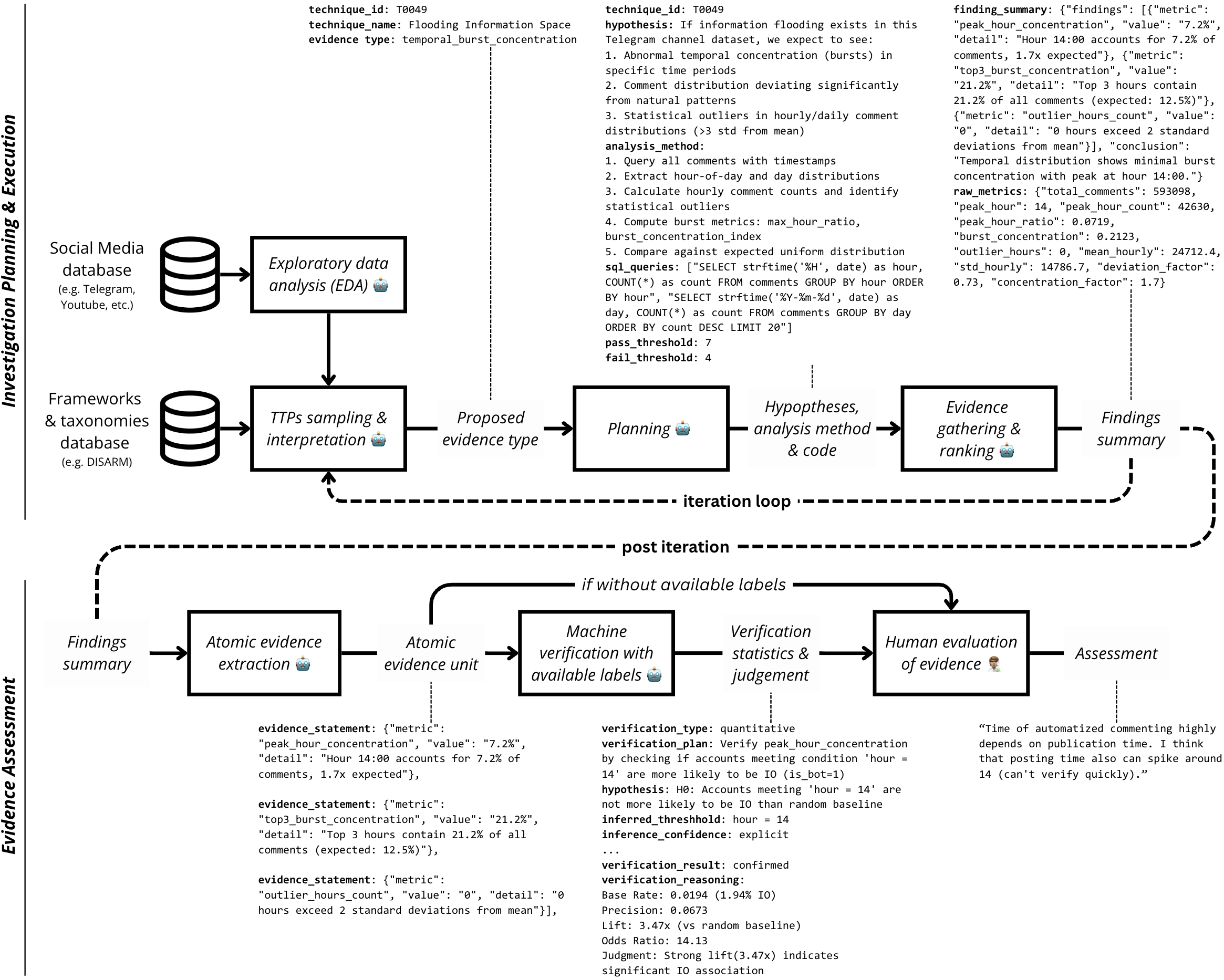}
    \caption{Workflow of the proposed agentic investigation and evidence assessment.
The workflow consists of two sequential stages. In the investigation planning and execution stage (upper panel), planning proposes the hypotheses, which are subsequently examined using exploratory data analysis (EDA) and TTPs sampling over social media data and framework-based taxonomies. Planning, hypothesis-driven analysis, and evidence gathering are connected through an iterative loop, producing structured findings summaries across iterations. In the evidence assessment stage (lower panel), findings are decomposed into atomic evidence units, evaluated through automated machine verification (when labels are available), and followed by a final human evaluation.}

    \label{fig:pipeline}
\end{figure*}

\subsubsection{{Russian campaign on Telegram for 2025 Moldova Election}}
This dataset captures coordinated and automated political commenting activities on Moldovan Telegram channels during a period of heightened electoral interference. The data were collected and verified by the OpenMinds research team as part of its investigation into Russian information operations targeting Moldova (see published report in \cite{openminds2025telegrambots}). Figure~\ref{fig:moldova_dataset} illustrates an instance of bot comments in original language and English translation.

The study is based on a list of Telegram channels obtained through a mapping of the Moldovan socio-political and informational Telegram ecosystem, from which a subset of channels relevant to political discourse was identified. From this mapped set, only channels with open comment sections were selected for analysis, resulting in a final sample of 88 political and news channels.

Bot identification focused exclusively on user-generated comments posted in these open comment sections between July 1 and September 15, 2025. During this period, 462 unique accounts that exhibited bot-like behavior were identified. These accounts collectively produced 62{,}229 comments on the selected channels. All identified accounts were manually verified and classified as bots following a structured review process.

Bot detection relied on a combined methodology that integrated algorithmic indicators and expert human judgment. Algorithmic signals included unusually high commenting volume, temporal regularity, repetitive posting patterns, and coordinated activity across multiple channels. These signals were subsequently assessed through manual inspection to exclude high-activity organic users and ensure that only coordinated or automated behavior was retained in the final data set.

Following the publication of the initial study, data collection was extended until the end of September 2025 to capture activity on election day. The final dataset comprises 595{,}814 total comments, produced by 30{,}297 unique accounts, of which 584 accounts were verified as bots. Bot accounts generated 86{,}448 comments, representing 14{.}5\% of all comments in the data set. The final released dataset therefore represents a high-confidence corpus of bot-driven political commenting activities on Moldovan Telegram, grounded in manual verification and consistent behavioral classification.


\subsection{{DISARM framework integration}}
The DISARM (Disinformation Analysis and Risk Management) framework provides a standardized taxonomy for characterizing information operations, organized hierarchically into tactics, techniques, and sub-techniques. Our system integrates DISARM at multiple levels:

\begin{itemize}
    \item \textbf{Technique selection}: The AI agent queries the DISARM database to identify applicable techniques
based on data characteristics and prior findings.
    \item \textbf{Finding attribution}: Each evidence gathering result is tagged with the specific DISARM technique ID (e.g. \texttt{T0049 flooding the information space}).
    \item \textbf{Evidence traceability}: Verified atomic evidence maintains links to both the source finding and the attributed DISARM technique.
\end{itemize}

\noindent This integration enables standardized reporting compatible with threat intelligence frameworks and facilitates cross-dataset comparison of IO tactics.

\subsection{{Workflow}}
Figure \ref{fig:pipeline} illustrates the end-to-end workflow of our pipeline that operationalizes the DISARM framework to perform automated analyses of social media content. It consists of the following key stages:

\subsubsection{{Exploratory data analysis (EDA)}}
The workflow begins with its first iteration by ingesting raw social media data into a SQL-queryable format. The first iteration performs an EDA that captures the dataset's statistics--sums, unique accounts, temporal range, and platform-specific metadata--without conducting anomaly detection. This deferred approach avoids premature attribution, allowing
DISARM alignment to be established only in the later technique-guided iterations.

\subsubsection{{TTPs sampling and interpretation}}
Beyond the first iteration of EDA, the AI agent selects DISARM techniques using an LLM-guided explore–exploit strategy inspired by the multi-armed bandit literature \cite{Katehakis1987}, rather than a classical statistical bandit algorithm.

\begin{itemize}
    \item \textbf{Explore}: Prioritize untested techniques to maximize coverage of DISARM taxonomies.
    \item \textbf{Exploit}: Deep-dive into techniques showing promising signals from prior
iterations.
    \item \textbf{Strategy adaptation}: The explore-exploit balance is dynamically adjusted based on cumulative investigation outcomes—shifting toward exploration after repeated failures and toward exploitation after discovering strong signals.
\end{itemize}

\noindent Selection criteria include: (1) new insight potential, (2) pattern matching with observed data characteristics, (3) operationalizability via SQL queries, and (4) specificity (sub-techniques preferred over parent techniques). Upon selecting a technique by its ID and description (e.g., \texttt{T0049 flooding the information space}), the agent proposes an evidence type (e.g., temporal burst concentration), which is further operationalized in the planning stage.  

\subsubsection{{Planning}}
Based on the selected DISARM technique and an LLM-interpreted evidence type, the AI agent elaborates the input into a set of investigation components, including:

\begin{itemize}
    \item Hypothesis (e.g.): 
    
    \texttt{If information flooding exists in this Telegram channel dataset, we expect to see: 1. Abnormal temporal concentration (bursts) in specific time periods  ... 3. Statistical outliers in hourly/daily comment distributions (>3 std from mean)}

    \item Analysis method (e.g.): 
    
    \texttt{1. Query all comments with timestamps; 2. Extract hour-of-day and day distributions; 3. Calculate hourly comment counts and identify statistical outliers; 4. Compute burst metrics max\_hour\_ratio, burst\_concentration\_index; 5. Compare against expected uniform distribution}

    \item SQL queries (e.g.): \begin{verbatim}
SELECT strftime('%H', date) AS hour,
       COUNT(*) AS count
FROM comments
GROUP BY hour
ORDER BY hour
...
\end{verbatim}
    \item Success criteria (e.g.): 
    
    \texttt{pass\_threshold: 7 fail\_threshold: 4}
\end{itemize}

\noindent These are used by the AI agent to initialize the acquisition of evidence from the target data set.

\subsubsection{{Evidence gathering and ranking}}
The AI agent executes the investigation plan by running generated SQL queries and Python scripts on the dataset to calculate the behavioral metrics. It then interprets the result in context—evaluating observed patterns against platforms, dataset characteristics, and prior findings—and assigns a signal strength score (0–10 scale) and status:

\begin{itemize}
    \item \texttt{PASS} (7.0–10.0): Strong evidence of the technique. 
    \item \texttt{INCONCLUSIVE} (4.0–6.9): Weak or ambiguous signal.
    \item \texttt{FAIL} (0.0–3.9): No evidence found.
\end{itemize}

\noindent All results are logged onto a structured database, enabling full traceability and downstream verification.

\subsubsection{{Atomic evidence extraction}}
Investigation findings often contain multiple claims within a single summary result. The extraction phase decomposes these into \textit{atomic evidence units}—discrete, independently verifiable claims that can be individually tested against ground truth data or annotations. Evidence types are categorized according to the nature of the claim (e.g., quantitative metrics, temporal patterns, content similarity, network structure) and rated by strength based on deviation from baseline.

\subsubsection{{Machine verification with available labels}}
Each atomic evidence is converted to a SQL detection condition and evaluated against available labels annotated through human-, machine-, or hybrid-workflows. The verification process consists of the following step:
\begin{enumerate}
    \item \textbf{Condition extraction}: Parse the natural language evidence into an executable SQL predicate.
    \item \textbf{Population application}: Identify all accounts that match the detection condition.
    \item \textbf{Confusion matrix}: Compare predictions against ground truth IO labels or annotations to compute TP, FP, FN, TN. 
    \item \textbf{Statistical testing}: Apply effect size calculation, Odds Ratio (OR), and significance testing (Fisher’s exact test).
\end{enumerate}

\noindent An evidence claim is marked \texttt{PASS} if all following three criteria are met: (1) the inference confidence is explicit (threshold clearly stated in the evidence), (2) the effect size meets the threshold (OR $\geq$ 3.0), and (3) the statistical significance (p $<$ 0.05).

\subsubsection{{Human evaluation of evidence}}
We complement statistical verification with human evaluation by IO domain practitioners, who qualitatively review the statistically verified evidence claims for both datasets. We report the qualitative evaluation findings in Section IV.

\subsection{{Pipeline implementation}}
The pipeline is implemented according to the following configuration:
\begin{itemize}
    \item \textbf{LLM model}: Claude Opus 4.5 (Anthropic), selected for its strong reasoning capabilities and extended context window.
    \item \textbf{Development environment}: Claude Code, Anthropic’s CLI-based AI agent for software development tasks.
    \item \textbf{Framework architecture}: Built on BMAD-METHOD, the system specifies agent workflows through markdown-based task definitions and YAML configurations. This natural language approach eliminates traditional programming frameworks, allowing intelligence analysts to customize detection workflows without coding expertise.
    \item \textbf{Data storage}: SQLite database for evidence gathering results and verification output, enabling SQL-based analysis and reproducibility.
    \item \textbf{Workflow parameters}: Maximum 15 iterations per dataset, signal strength threshold of 7.0 for \texttt{PASS} status.
    \item \textbf{Long-running execution}: To maintain instruction adherence across 15 iterations, the workflow employs subprocess relay architecture with persistent SQLite storage serving as cross-session memory. Each iteration executes as an independent subprocess while inheriting the full investigation history, preventing context drift and enabling consistent behavior throughout the investigation.
\end{itemize}
The complete workflow—from raw data ingestion through statistical verification—executes autonomously with human review only at the final stage.

\section{Results}
We execute 15 iterations per dataset: one initial exploratory data analysis followed by 14 iterative investigation rounds. Each round builds on prior findings through the full-history feedback mechanism, enabling progressive refinement of detection strategies. By design, each investigation round produces at most 3 atomic evidence claims, yielding a maximum of $14 \times 3 = 42$ evidence claims per dataset.

We evaluate evidence quality at two levels:
\begin{itemize}
    \item \textbf{Atomic evidence level} ($N=84$): Each evidence claim independently evaluated.
    \item \textbf{Technique detection level} ($N=28$): \texttt{PASS} if \textit{any} evidence in an iteration passes.
\end{itemize}

\subsection{Atomic Evidence Pass Rate}

\begin{table}[H]
\centering
\small
\begin{tabular}{@{}lrrrr@{}}
\toprule
Dataset & $N$ & Pass Rate & PASS & FAIL \\
\midrule
Russia/Telegram & 42 & \textbf{38.1\%} & 16 & 26 \\
China/X & 42 & \textbf{19.0\%} & 8 & 34 \\
\midrule
\textbf{Combined} & 84 & \textbf{28.6\%} & 24 & 60 \\
\bottomrule
\end{tabular}
\caption{Atomic evidence pass rate by dataset.}
\label{tab:atomic_pass_rate}
\end{table}

\subsection{Technique Pass Rate}

\begin{table}[h]
\centering
\small
\begin{tabular}{@{}lrrrr@{}}
\toprule
Dataset & $N$ & Pass Rate & PASS & FAIL \\
\midrule
Russia/Telegram & 14 & \textbf{64.3\%} & 9 & 5 \\
China/X & 14 & \textbf{35.7\%} & 5 & 9 \\
\midrule
\textbf{Combined} & 28 & \textbf{50.0\%} & 14 & 14 \\
\bottomrule
\end{tabular}
\caption{Technique pass rate by dataset.}
\label{tab:technique_detection}
\end{table}

\subsection{Human Evaluation}
\subsubsection{Chinese campaign on X}

The key finding is the detection of a suspicious burst in account creation dates that, after evaluation, we corroborate as being related to the creation of IO accounts. This pattern was previously observed in manual analyses of the full dataset. Most accounts created in August 2018 are IO accounts (precision = 0.97), demonstrating the agent-picked algorithm's ability to surface temporally clustered, coordinated account creation.

Additional signals refining this pattern—such as empty profile descriptions, campaign-related keywords, and interaction patterns—do not substantially improve detection, given the already high precision of the parent signal. However, they provide useful characterizations of subgroups of the IO accounts, including the characteristics of their activity and content.

The main signal absent from agent-generated evidence is temporal synchronization. For example, two users repeatedly retweet the same posts within a short time window (e.g., 10 minutes) over multiple days. This type of coordination is present in the dataset for retweets, replies, and especially mentions, but the current pipeline does not capture the temporal dimension.

\subsubsection{Russian campaign on Telegram}
The proposed technique helped identify a subset of bot accounts that had not been detected during the initial human-led investigation. Specifically, pattern-based analysis surfaced more than 30 additional accounts whose behavioral characteristics were consistent with coordinated or automated activity, demonstrating the value of the algorithm as a complementary detection tool.

It also captured several auxiliary metrics that aligned with patterns previously observed by human analysts. These included, among others, duplicated comments, shorter average comment length, and a lower mean number of reactions received by bot-generated comments. While such features proved useful in corroborating hypotheses regarding the artificial nature of borderline or ambiguous (“gray-zone”) accounts, they were insufficiently robust to serve as primary indicators for bot detection when considered in isolation.

At the same time, certain indicators employed during human-led identification did not emerge among the algorithmically selected features. Notably, these included measures such as consecutive days of sustained activity and the repeated dissemination of identical links to AI-generated video content produced as part of the information operation. This discrepancy highlights both the complementary strengths and the current limitations of automated approaches compared to expert-driven qualitative assessment.

\subsection{Cost \& Efficiency}
The workflow was executed using Claude Code. When running on Claude Opus 4.5, the process was deliberately split into multiple runs to prevent short-cutting iterative steps. In total, 15 iterations were performed, including an initial iteration dedicated to exploratory data analysis. The total API runtime was approximately 35 minutes, with a total cost of \$11.40 USD.

\section{Discussion}

\subsection{{Automated analyses vs. human analyses}}

At this stage, the output of the automated analysis should be understood as a structured set of techniques and patterns that can support bot identification, rather than as a fully autonomous detection system. CIB detection is typically based on a rule-based decision framework in which heterogeneous signals—both behavioral and content-based—carry different weights and are combined to classify individual accounts. In this study, the automated approach produced a consolidated table of TTPs and their operationalization within this specific dataset. However, the absence of an explicit decision tree or a weighted set of rules underscores the continued need for human judgment in the final classification process.

Not all proposed metrics proved equally effective in this dataset. Several thresholds required additional contextual constraints to be informative. For instance, message duplication is a well-established signal of coordinated activity, yet its practical use depends on factors such as message length, as short or emoji-based comments may be frequently repeated by organic users. Similarly, message length itself emerged as a potentially relevant signal but, as noted in the Results section, remains insufficient as a standalone indicator.

A focus on purely behavioral patterns brings both advantages and limitations. Its primary strength lies in its relative universality and reduced dependence on linguistic or political context. At the same time, this approach limits analytical resolution in several important respects. Behavioral signals alone do not allow researchers to distinguish between multiple bot farms operating simultaneously, to separate coordinated groups acting in opposition to one another, or to differentiate between bots and trolls (the latter referring to human-operated accounts engaged in coordinated or semi-coordinated manipulation). Although generative AI is blurring the boundary between these categories, in this dataset they remain associated with distinct content characteristics—automatically generated versus human-authored—which is a distinction that remains operationally relevant for designing effective countermeasures or statements on inauthenticity.

These limitations point to the potential value of incorporating a temporal dimension into automated detection. Longitudinal CIB operations evolve over time, as do the patterns through which they manifest. In this case, the same bot infrastructure has been observed operating in other contexts, including Russian-occupied territories of Ukraine \cite{dukach2024digital}, with partially overlapping accounts. Over the past year, this operation appears to have shifted from highly repetitive commenting strategies to more diverse AI-generated messages. Capturing such shifts through temporal modeling could improve long-term detection and make adaptation itself a detectable signal of coordinated influence.

Despite these constraints, human evaluation suggests that the techniques and patterns identified by agentic AI remain valuable as research aids. Rather than replacing expert analysis, they function as structured guides that highlight relevant signals and narrow the analytical focus, supporting a more consistent and transparent human‑led assessment of coordinated inauthentic behavior.

\subsection{{Limitations and future work}}
While the agentic system excels at recognizing behavioral patterns, it is currently limited in capturing temporal synchronicity and nuanced linguistic shifts. Future work should focus on modeling temporal evolution to detect and surface shifting campaign tactics, techniques, and procedures. To further increase the method’s generalizability, the pipeline should be evaluated across a broader range of social media platforms and FIMI datasets. Additionally, to increase its interpretive capacity beyond textual content and for context-dependent or specific settings (e.g., social media usage under kinetic conflict or occupation \cite{Wu2025DigitalFrontlines}), subsequent efforts will increase the grounding of AI agent workflow in human FIMI practitioners' operational context and expertise through a principled mixed methods approach as advocated by \cite{Lutz2026VisualIntegrity}.

\section{Conclusion}
This paper presents an agentic pipeline to operationalize the DISARM framework for augmenting FIMI investigations on social media. We demonstrate a novel approach in which autonomous AI agents collaboratively detect manipulative behaviors and map them to standardized TTPs.

Our evaluation on two real-world datasets yielded a 50\% technique pass rate across 14 autonomous iterations. In particular, the system successfully identified more than 30 previously undetected bot accounts in the Telegram dataset, demonstrating its ability to surface actionable intelligence beyond initial human benchmarks. These results confirm that autonomous agents can function as a cost-effective force multiplier, compressing the investigation timeline while ensuring outputs remain doctrinally aligned for interoperability. While human oversight remains essential for assessing intent, this work establishes a verifiable baseline for automated FIMI investigations in high-tempo hybrid defense environments.

\bibliographystyle{IEEEtran}
\bibliography{references}

\vspace{12pt}

\end{document}